# A NEW LEED INSTRUMENT FOR QUANTITATIVE SPOT PROFILE ANALYSIS

*U. Scheithauer, G. Meyer and M. Henzler*

*Institut für Festkörperphysik, Universität Hannover, Appelstr. 2*

## Keywords

*LEED, spot profile analysis, instrumental development, epitaxy of Pb on Cu (111)*

## Abstract

A new instrument for spot profile analysis of electron diffraction (SPA-LEED) has been set up. The instrument works either with a transparent phosphor screen for visual inspection of the pattern or in its main mode with a channeltron for the measurement of the intensity. The diffraction pattern is recorded with a fixed channeltron position by scanning the beam over the channeltron aperture using two sets of electrostatic deflection plates. The scanning range covers about 30°. The intensity may vary over five orders of magnitude. The SPA-LEED system was checked with the Si (111) 7 x 7 surface. A full width at half maximum of 0.3% of the normal reflex distance corresponding to a transfer width of 110 nm is reproducibly obtained. Under optimum conditions the transfer width rose up to about 200 nm. Initial high resolution measurements have been performed on the system Pb on Cu (111). The results demonstrate the possibilities of the new instrument for qualitative and quantitative analysis.

## 1. Introduction

Low Energy Electron Diffraction (LEED) is the most widely used technique for analysis of surface structures. The LEED pattern is monitored in nearly all UHV-systems to characterize on the spot the crystallinity or in general the reproducibility of the surface structure. For quantitative analysis during the last decades mostly the intensity of the diffracted spots has been used, which provides due to a well developed dynamic theory the position of the surface atoms within the unit mesh, provided the unit mesh is not too large and the surface is strictly periodic, that means defect-free (1, 2).

The last condition is usually checked by the more or less qualitative inspection, if the pattern is "good", that means sharp spots and low background. On the other hand there is strong demand to know surface defects qualitatively and quantitatively, since a lot of important surface processes are directly connected with deviations from periodicity like point defects, steps, domain boundaries or incomplete layers at least as an intermediate state. To study crystal growth, ordering of adsorbates, phase transitions or heterogeneous catalysis the knowledge of surface defects is very important. Therefore LEED has been used since long for that purpose by making use of the spot profile. It has been shown that for nearly perfect surfaces the evaluation of the spot profile is very much simplified by the validity of the kinematic approximation (3, 4). Unfortunately the usual instrumentation provides a very low resolution and dynamics because of its electron gun and screen display. The vidicon-based acquisition systems improve the speed of data collection, not the resolution, which is given by the minimum diameter of a diffraction spot obtained from a strictly periodic surface. A lot of instrumental improvements have been obtained in different laboratories, compromising between resolution, handling, speed and practicability (5-10). The new system presented here combines the advantages of a screen system (direct inspection on a backview window) with the advantages of a faraday cup system (high resolution) in one system. A high speed is obtained by a channeltron detector.

## 2. Design of the system

The whole system is mounted on a 203 mm o.d. conflat flange for replacement of a normal four grid LEED optics. Fig. 1 shows the principal elements of the SPA-LEED system. The system has a glass phosphor screen, which is observed from the backside. The screen gives a quick available overview of the reflex pattern. Due to a distance of nearly 240 mm between crystal and screen the spatial resolution is higher than in comparison with the normal optics so that more details are recognized. On the other hand the visible area of the reflex pattern is smaller. Behind holes in the screen the electron gun and the channeltron are mounted. In the main mode the intensity at a given position of the pattern is detected by the channeltron. Using two sets of electrostatic deflection plates the whole diffraction pattern is scanned over the channeltron aperture. A crystal lens focuses the electron source with an image ratio of 1:1 onto the channeltron aperture of 0.1 mm diameter.

## 3. Electrostatic deflection

The main concept of the system is to handle the reflex scanning with electrostatic deflection (5, 6). No mechanical movement of the sample or of the detector is needed during scanning, which would reduce accuracy and speed considerably. The principle of the electrostatic scanning is shown in fig. 2 in real space. Without voltages applied, in that picture the direct beam is reflected into the channeltron aperture for a crystal with its surface inclined by 4.6° relative to the axis of the system. The mechanical construction determines the angle between incoming beam $\underline{k}_0$ and outgoing beam $\underline{k}_1$. For electrostatic scanning the opposite deflection plates are supplied with equal voltages of contrary polarity. The ratio of voltages between the screen plates and the crystal plates is chosen close to -1 so that the position of the primary beam does not shift on the crystal. The angle between $\underline{k}_0$ and $\underline{k}_1$ remains constant during scanning. Therefore the absolute value of the scattering vector $K = |\underline{k}_1 - \underline{k}_0|$ stays constant, only its orientation with respect to the surface varies during scanning. Therefore a modified "Ewald" construction with the centre of the sphere at 000 as shown in fig. 3 for the Si (111) surface has to be used. Here the scattering vector $\underline{K}$ is the radius of the sphere instead of the incidental vector $\underline{k}_0$ in the usual "Ewald" construction. Due to the doubled radius the scanning range in reciprocal space is doubled for a given energy and given deflection angle of the beam.

## 4. Electron gun

The electron gun supplies a current up to $10^{-6}$ A for the screen display although a fraction of it would be sufficient. If working with the channeltron a current of $10^{-10}$ A to $5*10^{-8}$ A is needed. For that low current the electron source diameter is less than 0.1 mm down to an electron energy of about 30 eV. The electron gun uses a commercial direct heated tungsten filament with a tip. For that small diameter special efforts are needed with adjustment and design of cathode, Wehnelt and anode region (11). The tip of the cathode is adjusted with a screw driven tripod to a precision of less than 0.1 mm with respect to the axis of the gun and the distance to the Wehnelt electrode.

## 5. Data acquisition and control of the experiment

Voltages may be applied to the deflection plates under computer control. At fixed primary electron energy the count rate for any scan position is recorded. The computer generates one-dimensional line scans or two-dimensional area scans. In the case of area scans the intensity can be presented as contour level or surface plot with different magnifications (figs. 5, 7, 8). Dependent on the scale the area scan gives either an overview of the whole Brillouin zone or detailed information of one or a few spots only. Mainly for area scans the aspect of measurement time becomes important. The total measurement time depends on the total number of channels and for each channel on the desired accuracy. With a number of $10^4$ channels an area of 100x100 points may be recorded. Linear scans of at least 1000 channels are used for accurate measurements throughout the Brillouin zone to enable reliable quantitative evaluation (figs. 4, 6, 10).

## 6. Test of the system

The SPA-LEED was checked with a Si (111) 7 x 7 surface. There are two points characterizing the system, the FWHM (full width at half maximum) and the area of the reciprocal lattice that can be recorded by electrostatic deflection.

Fig: 4 shows in the upper part a linear scan through the (00) reflex and one of the first order superstructure reflexes and in the lower part two expanded scans through the (00) reflex in normal directions. The FWHM is better than 0.17% of the normal reflex distance, which corresponds to a transfer width of 210 nm (12). If this half width is measured with a high accuracy, distances of more than 500 nm are resolved (13). Since all measurements need a crystal for the reflection of the beam only an ideal crystal will enable the measurement of the instrumental function. Here therefore a silicon crystal had to be oriented, cut and polished to better than 0.05°. Careful cleaning and annealing is needed. The selection of energy of 96 eV avoids broadening due to any remainder of atomic steps. Both mechanical vibrations and any ripple of the deflection voltages (< 0.02%) has to be kept at a minimum. Fig. 5 shows an area scan of the reciprocal lattice over two Brillouin zones of the Si (111) 7 x 7 surface. The intensity is plotted versus the deflection voltages in a contour level plot. To accentuate the ring of the first normal reflexes they are connected by lines. The plot shows a scanning range of about 30°. The range can be extended by mechanical sample rotation. For large angles of deflection small distortions and a change of the deflection sensitivity up to 10% are present, which may be cancelled by a rescaling with the computer.

Fig. 6 shows a linear scan through the (00) reflex and two first order superstructure reflexes. First a scan with nearly $4.4 * 10^{-4}$ counts per second in the specular beam was recorded. Then the primary current was increased by a factor of 150 and a second scan was taken to measure the background with high accuracy. To avoid damage of the channeltron due to high counting rates the sweep leaps over the high count region of the (00) reflex. Near the (00) beam a weak spot becomes clearly detectable. An energy variation of the primary electron beam pointed out, that this intensity belongs to a facet of the surface with the orientation (335). Simple intensity ratio of the facet to the (00) beam points to a fraction of a few percent of the surface occupied by the facet. The increased half width of the facet beam yields a width of some 10 nm.

Fig. 7 shows a surface plot of the Si (111) 7 x 7 structure with the (00) reflex and the first ring of superstructure reflexes. In fig. 8 the same set of data as in fig. 7 is presented as a contour level plot. Since the contour levels are chosen below 1.5% o f the maximum count rate of the record a modulation of the background between the spots is visible, which may be a hint to an incomplete annealing of the surface.

## 7. Pb on Cu (111)

The system Pb on Cu (111) had already been studied with a conventional four grid optics (14, 15). Two structures had been observed. Up to a coverage of $\theta$ = 1 ML, calibrated against the Pb bulk distance, the Pb layer is always seen with nearly the Pb bulk distances. If additional Pb up to 1.07 ML is evaporated on top of this layer at room temperature, the Pb layer changes into a commensurate 4 x 4 structure. Fig. 9 summarizes this observed LEED pattern. High resolution measurements with the new SPA-LEED instrument have been performed to study the transition region between the two structures in detail. Linear scans within the marked region (s. fig. 10) were taken. The results are shown in fig. 10. Additional deposition of Pb at room temperature on top of a crystal with 1 ML Pb gives a continuous compression up to 3% at $\theta$ = 1.07 ML. This compressed structure is a commensurate 4 x 4 structure. During heating to at least T = 650 K the original Pb layer is reproduced again. Here, however, not a continuous change of the lattice constant, rather a growth of domains with bulk distances within the compressed layer is observed. These differences at intermediate coverages are hardly observable with a conventional LEED optics.

## 8. Conclusions

The facilities of the new SPA-LEED instrument have been pointed out. The instrument is characterized by its high resolution, a transfer width of better than 200 nm at optimum and a high dynamic range due to channeltron detection of the intensity. The deflection pattern is obtained by electrostatic deflection, no mechanical movement of detector or crystal during scanning is needed. For a quick overview a phosphor screen is incorporated. The system may replace a conventional system due to its convenient operation and will provide substantial additional information due to the precision spot profile measurements.

### Acknowledgement

The investigations have been supported by the Deutsche Forschungsgemeinschaft. The silicon single crystals have been kindly provided by Wacker-Chemitronic, Burghausen. The instrument was manufactured by Leybold Heraeus, Köln.

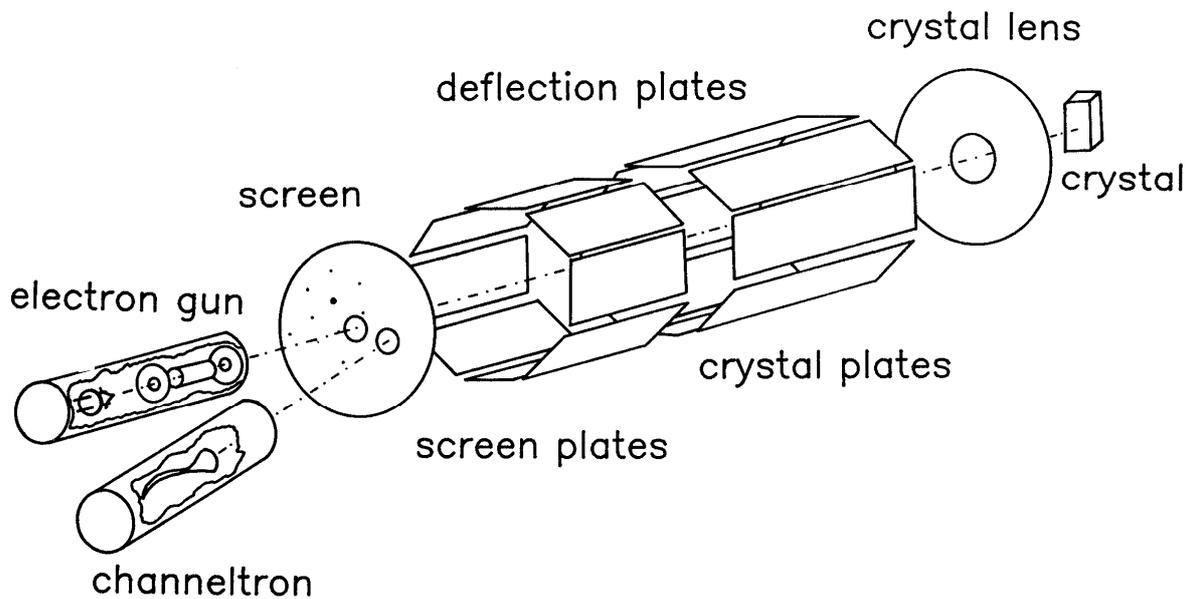

Fig. 1: Schematic set up of the SPA-LEED system

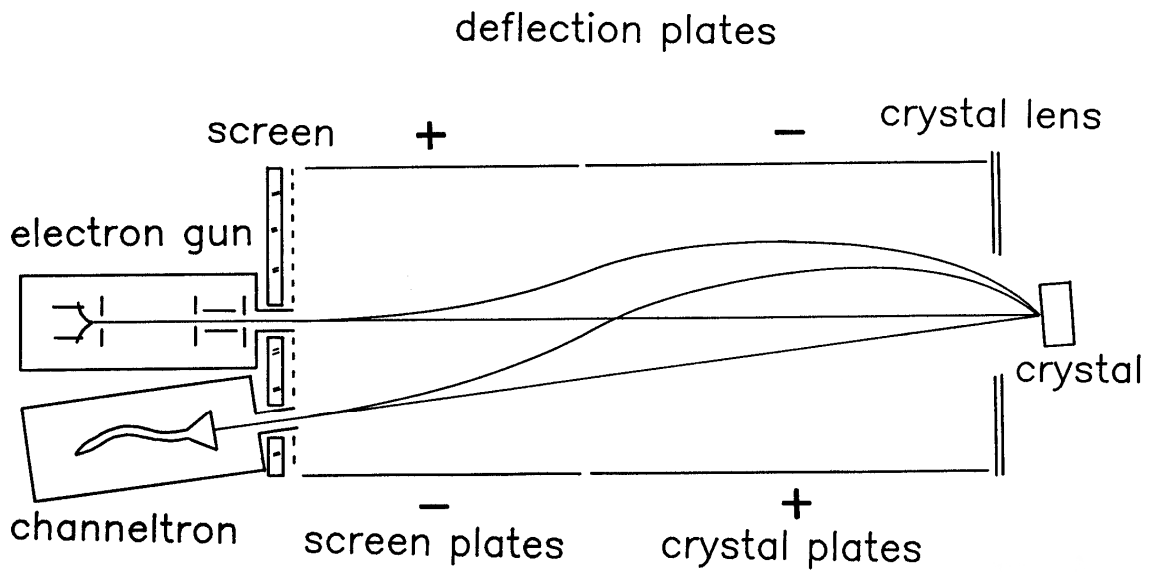

Fig. 2: Electrostatic deflection: Without deflection voltages the specular beam is detected by the channeltron. While applying voltages at the plates, both the incoming and outgoing beams are deflected simultaneously in the same way as shown in the figure. In this way any scattering vector may be obtained.

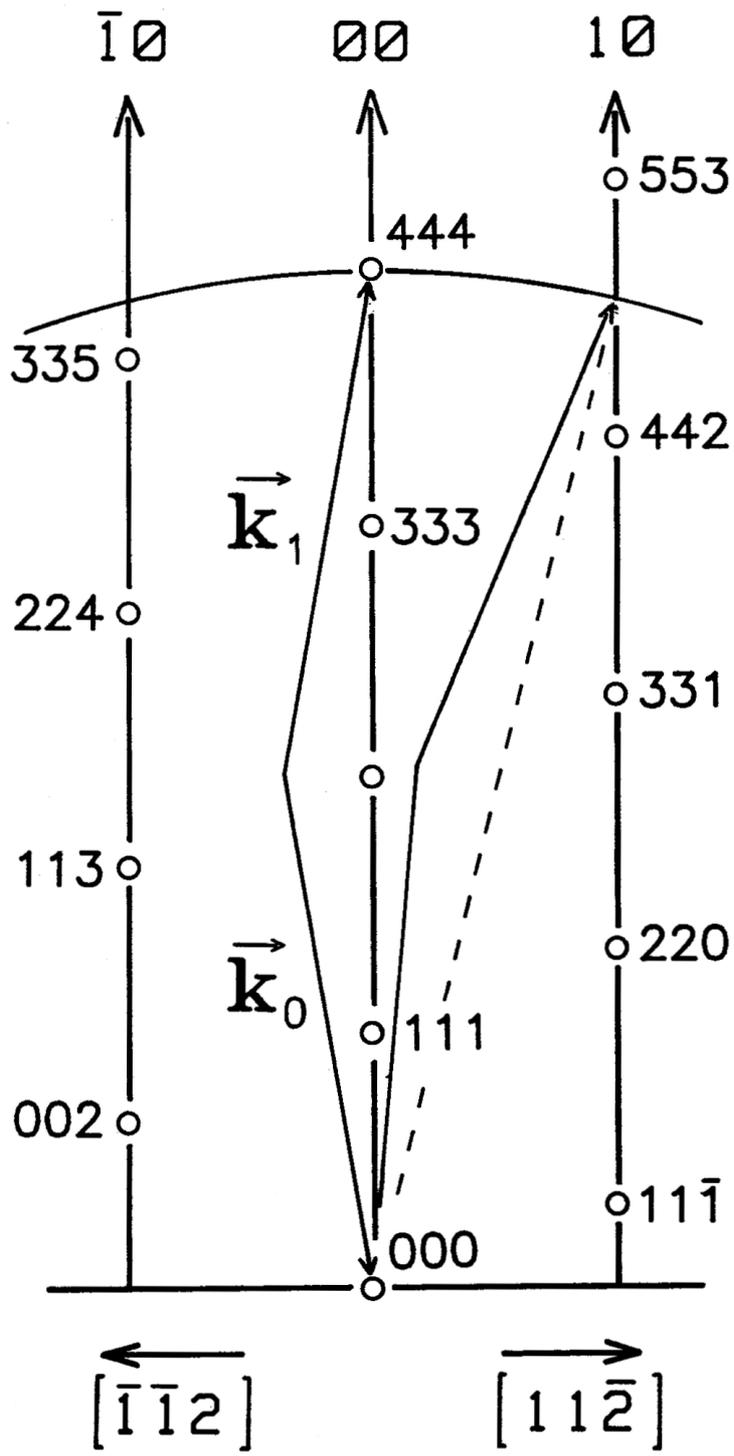

Fig. 3: The reciprocal space for the Si (111) surface with a modified "Ewald" construction. The sphere has a radius of $K = |\underline{k}_1 - \underline{k}_0|$ and its centre is at 000. $\underline{k}_1$ and $\underline{k}_0$ are drawn for two different scattering conditions.

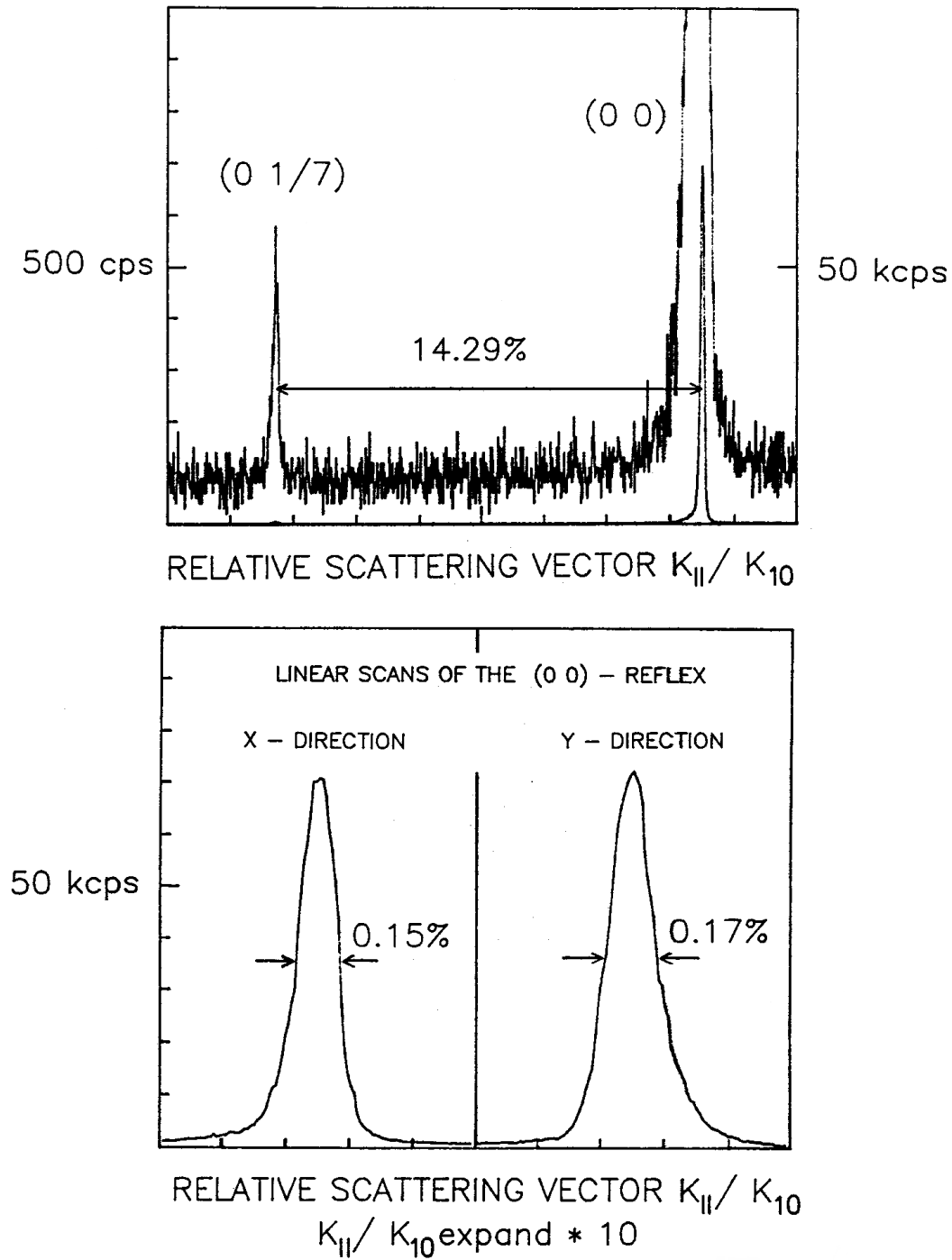

Fig. 4: Spot profile of the (00) reflex showing the resolution of the system. Si (111) 7 x 7, energy: 96 eV
Upper part: scan through the (00) reflex and one of the first order superstructure reflexes
Lower part: expanded scans in x and y direction through the (00) reflex

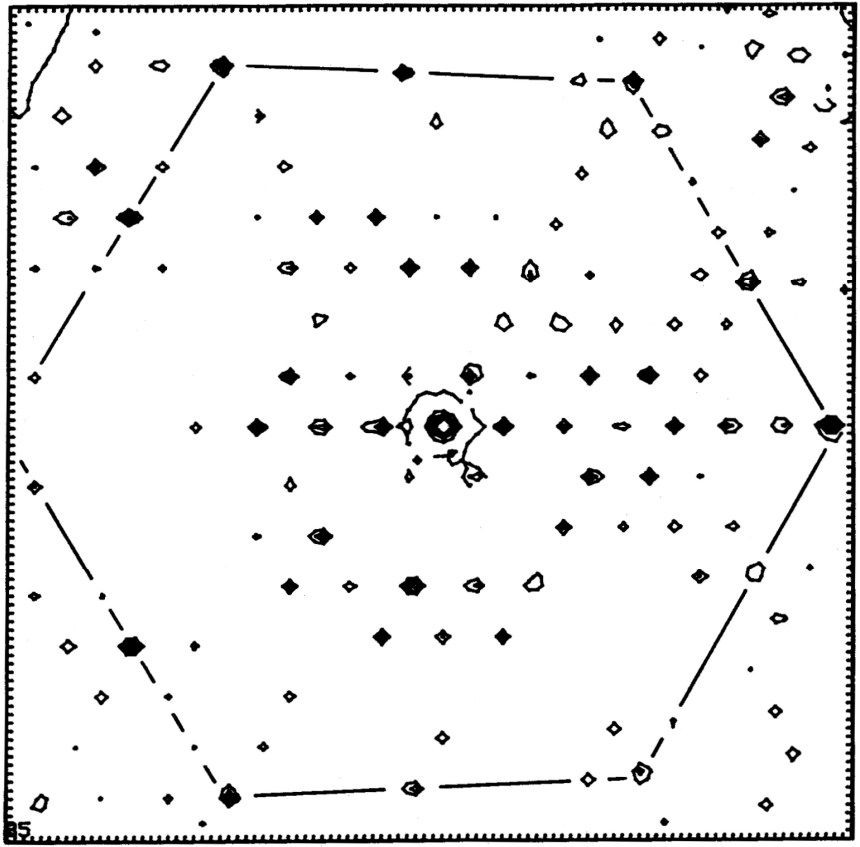

Fig. 5: Scan over two Brillouin zones of the Si (111) 7x7 surface by electrostatic deflection. The first order normal reflexes are connected by lines. Energy: 126 eV

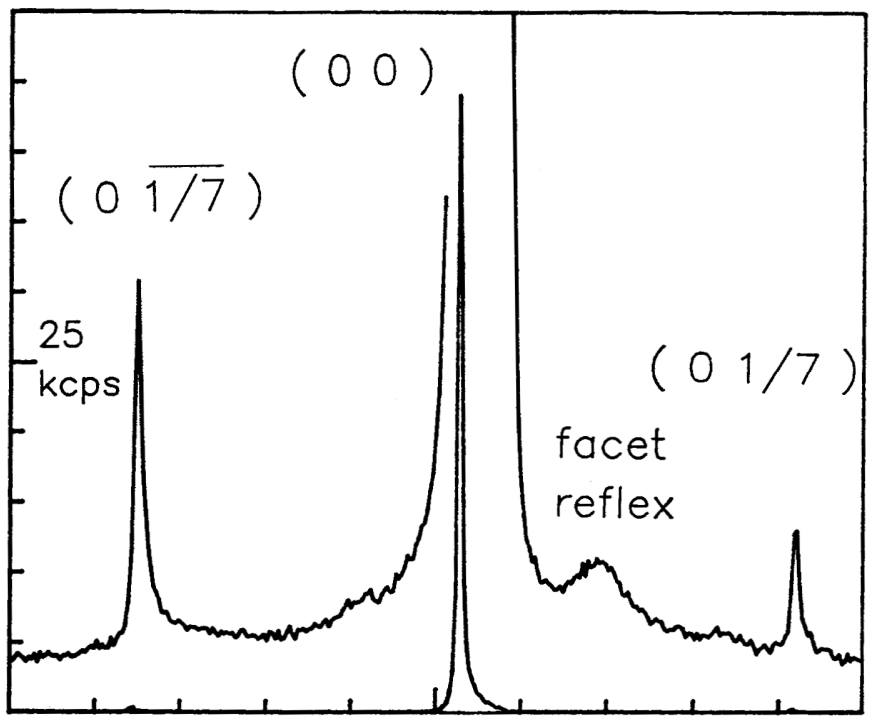

Fig. 6: Si (111) 7 x 7, "background" scan through the (00) reflex and two superstructure reflexes showing a facet reflex of (335) orientation, energy: 126 eV

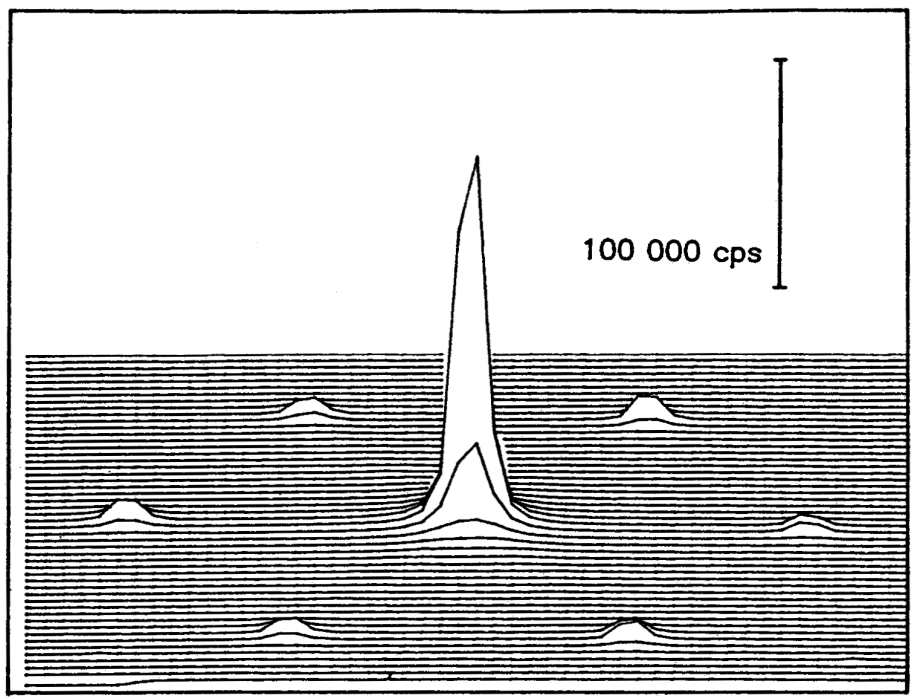

Fig. 7: Si (111) 7 x 7, surface plot of (00) reflex and ring of first superstructure reflexes, energy: 74.5 eV

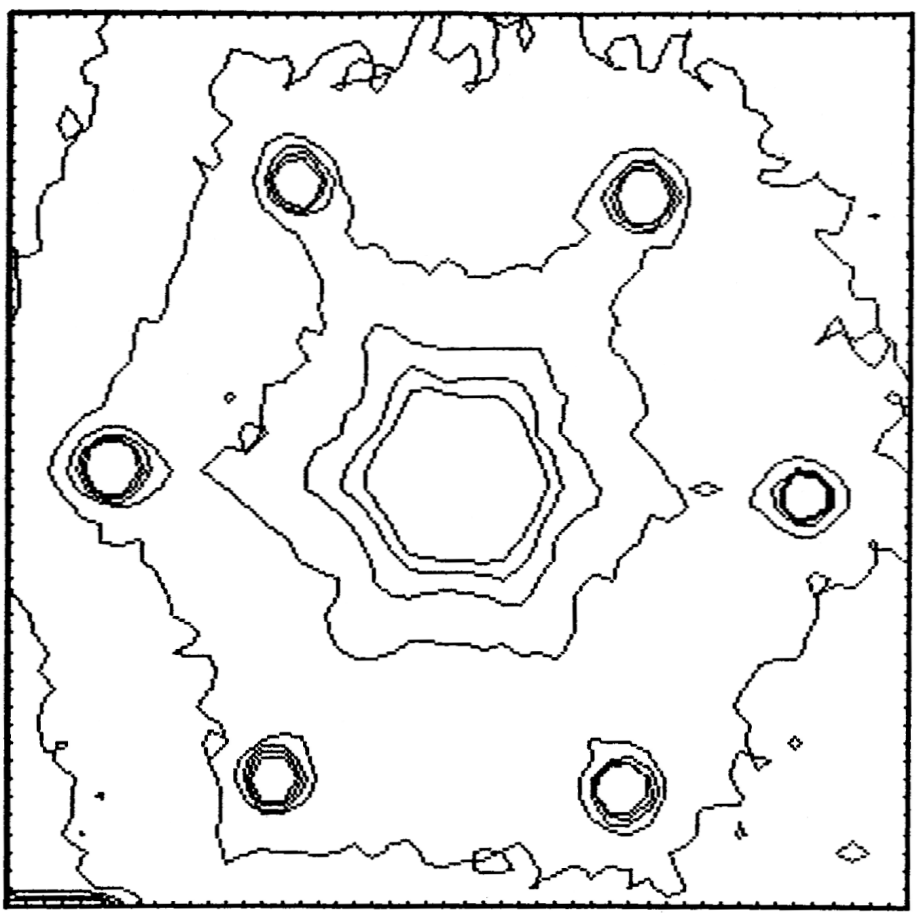

Fig. 8: Si (111) 7 x 7, same set of data as in fig. 7, contour level plot of (00) reflex and ring of first superstructure reflexes, energy: 74.5 eV

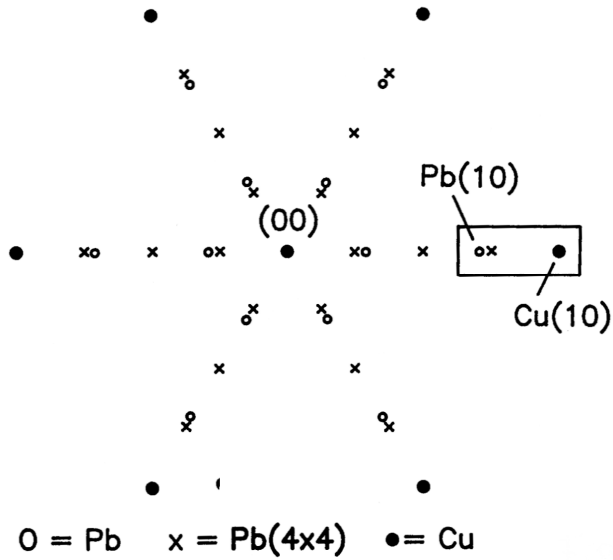

Fig. 9: Schematic LEED pattern of the Pb 4 x 4 and the Pb structure on Cu (111). The extra spots of the Pb 4 x 4 structure are shown only on the lines between the Cu spots. The rectangle marks the region were the linear scans, shown in fig. 10, were taken.

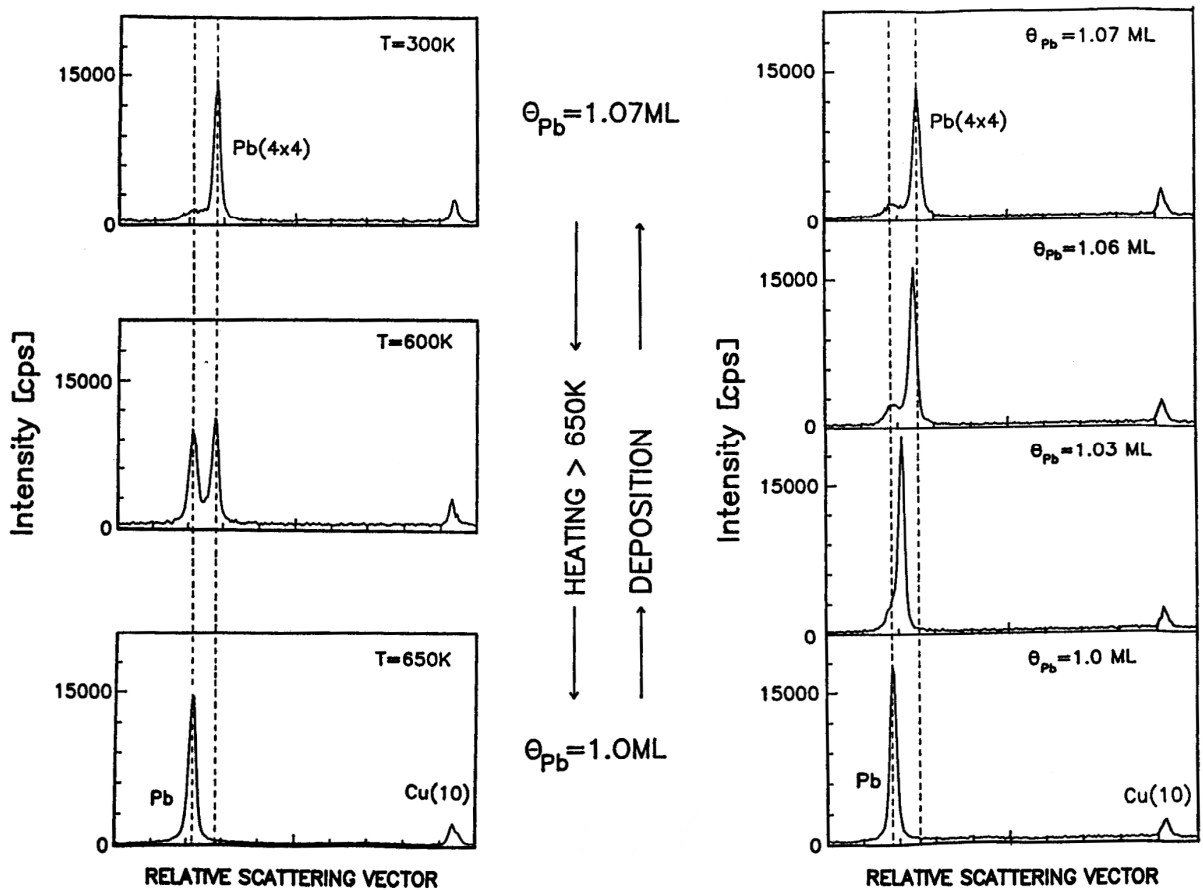

Fig.10: Right: With Pb deposition a shift of the Pb spot, that means a continuous compression of the Pb layer, is shown.
Left: The conversion to the uncompressed layer with definite heating periods is shown. The appearance of double spots is explained by the existence of domains of both structures. (Scans are taken at 300 K).